\documentclass{jkas}

\def\year{2019} 
\def\month{????} 
\def\volume{??} 
\def\issue{??} 
\def\beginpage{1} 
\def\endpage{99} 
\date{Received ; accepted }

\newcommand*{\teff}{$T_{\rm eff}$}
\newcommand*{\logg}{$\log~g$}
\newcommand*{\feh}{[Fe/H]}
\newcommand*{\kms}{km s$^{-1}$}
\newcommand*{\zmax}{$Z_{\rm max}$}

\newcommand*{\gaia}{$Gaia$}

\usepackage{color}
\usepackage{epsfig}
\usepackage{graphicx}
\usepackage{natbib}
\usepackage{hyperref}

\hypersetup{
    bookmarks=true,         
    unicode=false,          
    pdftoolbar=true,        
    pdfmenubar=true,        
    pdffitwindow=true,     
    pdfstartview={FitH},    
    pdftitle={},    
    pdfsubject={Astronomy},   
    pdfcreator={dvipdf},   
    pdfproducer={dvipdf}, 
    pdfnewwindow=true,      
    colorlinks=true,       
    linkcolor=red,          
    citecolor=blue,        
    filecolor=magenta,      
    urlcolor=cyan,           
    breaklinks=true,
    linktocpage
}

\title{
Origin and Status of Low-Mass Candidate Hypervelocity Stars
}

\author[1,2]{Bum-Suk Yeom}
\author[1]{Young Sun Lee}
\author[1]{Jae-Rim Koo}
\author[3]{Timothy C. Beers}
\author[1]{Young Kwang Kim}

\affil[1]{Department of Astronomy and Space Science, Chungnam National University, Daejeon 34134, Korea}
\affil[2]{Jeollabukdo Institute of Science Education, Iksan 54549, Korea}
\affil[3]{Department of Physics and JINA Center for the Evolution of the Elements, University
                 of Notre Dame, IN 46556, USA}

\def\corrauthor{
Young Sun Lee (youngsun@cnu.ac.kr)
}

\def\runningauthor{
Bum-Suk Yeom
}

\def\runningtitle{
Kinematic and Chemical Properties of Hypervelocity Stars
}

\def\keywords{
Method: data analysis -- technique: spectroscopy --
Galaxy: disk -- stars: abundances
}

\def\abstracttext{
We present an analysis of the chemical abundances and kinematics of six
low-mass dwarf stars, previously claimed to be candidate hypervelocity
stars (HVSs). We obtained moderate-resolution ($R\sim$ 6000) spectra of
these stars to estimate their abundances of several chemical elements
(Mg, Si, Ca, Ti, Cr, Fe, and Ni), and derived their space velocity
components and orbital parameters using proper motions from \gaia\ Data
Release 2. All six stars are shown to be bound to the Milky Way, and in
fact are not even considered high-velocity stars with respect to the
Galactic rest frame. Nevertheless, we attempt to characterize their
parent Galactic stellar components by simultaneously comparing their
abundance patterns and orbital parameters with those expected from
various Galactic stellar components. We find that two of our program
stars are typical disk stars. For four of the program stars, even 
though their kinematic probabilistic membership assignment suggests 
membership in the Galactic disk, based on their distinct orbital properties 
and chemical characteristics, we cannot rule out the exotic origins 
of these objects, as follows. Two stars appear to be runaway stars from the Galactic 
disk. One program star has possibly been accreted from a disrupted dwarf 
galaxy or dynamically heated from a birthplace in the Galactic bulge. The 
last object may be either a runaway disk star or has been dynamically heated.
Spectroscopic follow-up observations with higher resolution for these curious 
objects will provide a better understanding of their origin.}

\begin{document}
\jkashead 


\section{Introduction} \label{intro}

Hypervelocity stars (HVSs) are unbound and rare fast-moving objects in
the Galactic halo, possessing space velocities that exceed the Galactic
escape speed. The first HVS was discovered from a radial velocity
survey of faint blue horizontal branch stars \citep{brown2005}. It is a
3 $M_\odot$ main-sequence B-type star moving with a Galactic rest-frame
velocity of about 700 \kms\ at a distance of about 100 kpc. Since then,
about 20 B-type HVS candidates have been discovered in the Galactic halo
\citep{brown2015}. 

These intriguing objects are believed to originate from the so-called
``Hills mechanism'', which is associated with the supermassive black
hole (SMBH) at the Galactic Center. This theory suggests that, in the
case of a binary system interacting with the SMBH, the SMBH can destroy
the binary system and eject one of its stars, attaining speeds up to
$\sim$ 1000 \kms\ \citep{hills1988, yu2003}. It is known that this
mechanism can also produce HVSs bound to the Milky Way (MW)
\citep{bromley2009, brown2014}. 

In addition to the HVSs, there are other types of fast-moving stars,
referred to as ``runaway stars'' among O- and B-type stars, which have peculiar
velocities larger than 40 \kms\ \citep[e.g.,][]{gies1987, stone1991,
tetzlaff2011}. Several scenarios have been proposed to explain the
runaway stars. The binary ejection mechanism \citep{blaauw1961,
tauris1998,tauris2015} postulates that these objects could be formed in
a binary system and ``released'' out of their system by the explosive death
of their companion in the Galactic disk. The dynamical ejection
mechanism proposed by \citet{poveda1967} assumes that a star can be
ejected by multi-body interactions in a high-density environment such as
star clusters. Another explanation to account for these stars is the ejection from
a star-forming galaxy such as the Large Magellanic Cloud (LMC)
\citep{boubert2016}. An alternative theory suggests that these objects
are the members of a tidally disrupted dwarf galaxy \citep{abadi2009}.
Even though there exist many scenarios to explain the HVSs and runaway
stars, full understanding of their origin remains elusive. 

In spite of the ambiguity of their origin, high-velocity stars have
received attention because they can provide a means for measuring the
local escape velocity at a given distance from the Galactic Center,
which can in turn constrain the total mass of the MW, still uncertain by
more than a factor of two \citep{xue2008,watkins2010}. Even though the
origin of these objects is uncertain, we can utilize their general
characteristics to infer where they originated. What HVSs have in common
is that they are young, massive main-sequence stars typically found at
present distances beyond 50 kpc from the Sun \citep[See][and references
therein]{brown2015AR}. Therefore, we can hypothesize that the
high-velocity stars may originate from star-forming regions in the disk,
bulge, or dwarf satellites of the MW.

Although most of the currently known HVSs and runaway stars are early
type (high-mass) main-sequence stars, one might expect that the proposed
ejection mechanisms could work for any stellar type, leading to the
prospect of identifying low-mass HVSs \citep{kollmeier2007}. For this
reason, various efforts to search for such stars have been carried out
from the stellar database constructed by large spectroscopic surveys
such as Sloan Digital Sky Survey \citep[SDSS;][]{york2000} and Large
Sky Area Multi-Object Fibre Spectroscopic Telescope \citep[LAMOST;
][]{cui2012}.

Indeed, \citet{kollmeier2009} and
\citet{li2012} identified 6 and 13 HVS candidates, respectively, from
SDSS. In addition, \citet{zhong2014} reported 28 HVS candidates, 17 of
which are F-, G-, and K-type dwarf stars. \citet{li2015} reported
another 19 low-mass HVS candidates from LAMOST. \citet[][hereafter
Pal14]{palladino2014} also reported the discovery of 20 low-mass G-, and
K-type HVS candidates from Sloan Extension for Galactic
Understanding and Exploration \citep[SEGUE; ][]{yanny2009}. 
One interesting aspect of many of these low-mass candidates is
that they appeared to be associated with birth in the Galactic disk,
rather than the Galactic Center, as is the case for the high-mass HVSs. 

Previous studies of low-mass HVSs or runaway star candidates were
carried out exclusively on the basis of derived stellar kinematics,
since proper motion information, when combined with an observed radial
velocity and distance estimate, provides the full space velocity of a
star. In keeping with this, it is interesting to note that HVSs thought
to have a Galactic Center origin (e.g., the high-mass HVSs) are often
well-separated in their proper motions from likely disk origins, which
is not the case for the low-mass runaway-star candidates. For this
reason, one might consider use of their detailed chemical-abundance
patterns in order to identify the possible birthplaces of the
runaway-star candidates -- bulge, disk, or halo, and thereby constrain
the possible ejection mechanisms of such stars.

We note here that a number of recent studies claim that most of the low-mass
HVS candidates and runaway stars identified thus far are bound to the
MW. For example, \citet{ziegerer2015} have re-calculated the proper
motions for the 14 HVS candidates that Pal14 reported, using images from
SDSS\footnote{\url{http://skyserver.sdss3.org/public/en/tools/chart/navi.aspx}},
Digitized Sky Survey\footnote{\url{http://archive.stsci.edu/cgi-bin/dss_plate_finder}}
(DSS), and UKIDSS\footnote{\url{http://www.ukidss.org/}}
\citep{lawrence2007}. They found that the newly measured proper motions
are much smaller than the ones used by Pal14, and, as a result, all of
their HVS candidates are bound to the MW under three different
potentials for the MW. More recently, \citet{boubert2018} reported using
\gaia\ Data Release 2 \citep[DR2;][]{gaia2018} proper motions and
radial velocities that all late-type stars, which have been claimed to
be HVSs previously are likely to be bound to the MW, except one object
(LAMOST J115209.12$+$120258.0). 

Prior to clarification on the proper motions for low-mass HVS candidates
in the Pal14 sample, we carried out follow-up spectroscopic observations and obtained 
medium-resolution ($R$ = 6000) spectra for six of them, in order to study
their chemical abundance patterns. As we report in this paper, we make
use of {\it chemical tagging} \citep{freeman2002}, in an attempt to
understand their characteristics and likely parent populations of their
birthplaces. This approach has already proven to be useful to constrain
the origin of HVSs or runaway stars \citep{hawkins2018}. 
In addition, we make use of the greatly improved proper motion information
from {\it Gaia} DR2 to carry out a kinematics analysis of
our program objects. Although we confirm that none of our program stars
unbound, and thus are no longer viable HVS
candidates, for  simplicity we refer to them as HVS
candidates through this paper. This paper is organized as follows. The
spectroscopic observations and reduction of the six low-mass candidate HVSs
are described in Section~\ref{sec2}. In Section~\ref{sec3}, we determine
the stellar parameters and chemical abundances for our program stars. In
Section~\ref{sec4}, we present results of the analysis of the
chemical and kinematic properties of our objects. Section~\ref{sec5}
discusses the characteristics and possible origin of each HVS candidate.
A summary of our results, and brief conclusions are provided in Section~\ref{sec6}.

\begin{table}[t]
\caption{Details of spectroscopic observations \label{tab1}}
\centering
\resizebox{\columnwidth}{!}{
\begin{tabular}{lccccrc}
\toprule
Object & Magnitude  & Date (UT)   & Exposure  &  $N_{\rm{obs}}$ & S/N & Pal14~ID\\
       &       &             & Time (s)  &                 &  &        \\
\midrule
{\sc program stars:} & $r_{\rm{0}}$ & &  &   &  & \\
J113102.87+665751.1 & 16.15 & 2014.01.05/12 & 3600        &  5 & 49 & 4 \\
J064337.13+291410.0 & 18.01 & 2014.03.25    & 3600        &  3 & 30 & 7 \\
J172630.60+075544.0 & 18.46 & 2014.05.03    & 3600, 4500  &  3 & 30 & 10 \\
J095816.39+005224.4 & 17.51 & 2014.01.12    & 3600        &  3 & 40 & 15 \\
J074728.84+185520.4 & 17.81 & 2014.03.25/30 & 4500, 4000  &  2 & 29 & 16 \\
J064257.02+371604.2 & 16.87 & 2014.03.30    & 3600        &  3 & 37 & 17 \\ \addlinespace
{\sc standard stars:} & $V$ &  &  &  \\
HIP 33582   & 9.01  & 2014.01.05   & 1, 4, 8, 30     &  4 &\\
HIP 59330   & 8.59 & 2014.01.05   & 1, 3, 10, 15    &  4 &\\
HIP 44075   & 5.86 & 2014.01.12   & 1, 2, 3, 5      &  4 &\\
HIP 60551   & 8.00 & 2014.01.12   & 2, 3, 5, 10     &  4 &\\
HIP 64426   & 7.29 & 2014.01.12   & 1, 2, 3, 5      &  4 &\\
HR 2721     & 5.58 & 2014.03.25   & 1, 2, 5         &  3 &\\
HR 2233     & 5.65 & 2014.03.30   & 1, 1, 1, 1, 5   &  5 &\\
HIP 62882   & 8.38 & 2014.05.03   & 1, 3, 5, 10, 20 &  6 &\\
\bottomrule
\end{tabular}
}
\tabnote{$N_{\rm obs}$ is the number of observations. Pal14 ID is the ID of a star used in Pal14.
$r_{\rm{0}}$ indicates the reddening-corrected $r$ magnitude. S/N is the signal-to-noise 
ratio of the co-added spectrum of each star after data reduction. The S/N range of 
the standard stars can be seen in Figure \ref{figure3}.} 
\end{table}

\begin{figure} [t]
\includegraphics[width=\columnwidth]{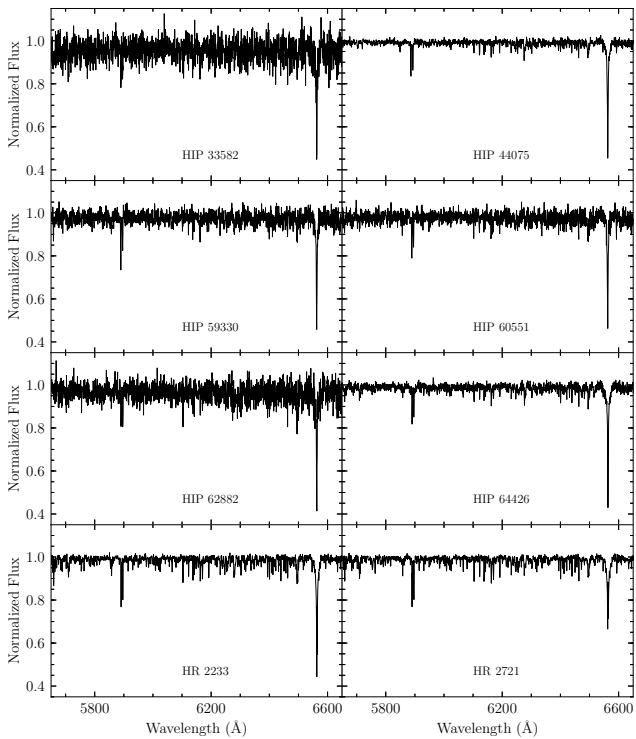}
\caption{Spectra of standard stars in the wavelength range 
used to derive stellar parameters and chemical abundances.}
\label{figure1}
\end{figure}

\begin{figure} [t]
\includegraphics[width=\columnwidth]{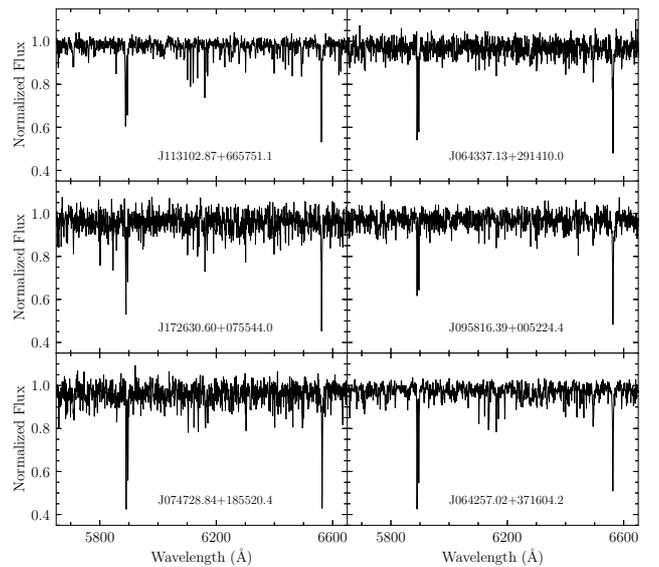}
\caption{Same as in Figure~\ref{figure1}, but for our program stars.}
\label{figure2}
\end{figure}

\section{Spectroscopic Observations and Reduction} \label{sec2}

Among the 20 HVS candidates reported by Pal14, we obtained spectroscopy
for six stars during the period between January and May, 2014, including
two stars that were not studied by \citet{ziegerer2015}. The spectra
were obtained with the Dual Imaging Spectrograph (DIS) on the Apache
Point Observatory 3.5 m telescope in New Mexico. We selected the
grating combination B1200/R1200, which has wavelength coverage of
4200 -- 5400\,{\AA} and 5600 -- 6700\,{\AA} in the blue and red channels,
respectively, along with a 1.5 arcsec slit, a yielding spectral
resolving power of $R$ = 6000, sufficient to perform a chemical
abundance analysis for individual $\alpha$-elements (Mg, Ca, Si, and Ti)
and iron-group elements (Cr, Fe, and Ni).

In addition, we observed eight well-studied disk stars as comparison
stars, to verify the accuracy of our derived stellar
parameters ($T_{\rm eff}$, $\log~g$, and [Fe/H]) and chemical abundances
for our program stars. For the comparison stars, we took several
exposures to obtain the spectra at various signal-to-noise (S/N) ratios,
enabling an evaluation of the effect of S/N on the estimated stellar
parameters and chemical abundances of our target stars. As our targets
are mostly faint ($r_{\rm 0} >$ 16.0), we took one or two exposures between 60 and 
75 minutes each. Table~\ref{tab1} lists details of the observations. 

We followed the standard spectroscopic reduction steps such as aperture
extraction, wavelength calibration, and continuum normalization using
IRAF\footnote{IRAF is distributed by the National Optical Astronomy
Observatory, which is operated by the Association of Universities for
Research in Astronomy, Inc., under cooperative agreement with the
National Science Foundation.}. Radial velocities were measured using the
cross-correlation function by applying \emph{xcsao} task of the
\emph{rvsao} package \citep{kurtz1998} in IRAF. In that process, we
used a synthetic model spectrum of $T_{\rm{eff}} = 6000$ K and log~$g$ =
4.0 as a template. We also considered the night-sky emission lines to
check for any additional instrumental shifts. The spectra of each star
were co-added after application of radial velocity corrections, to
obtain a final spectrum with higher S/N, typically S/N $\sim$ 30 -- 50. 
The spectra of our standard stars and program stars are shown in Figures 
\ref{figure1} and \ref{figure2}, respectively.

\begin{figure} [t]
\includegraphics[width=\columnwidth]{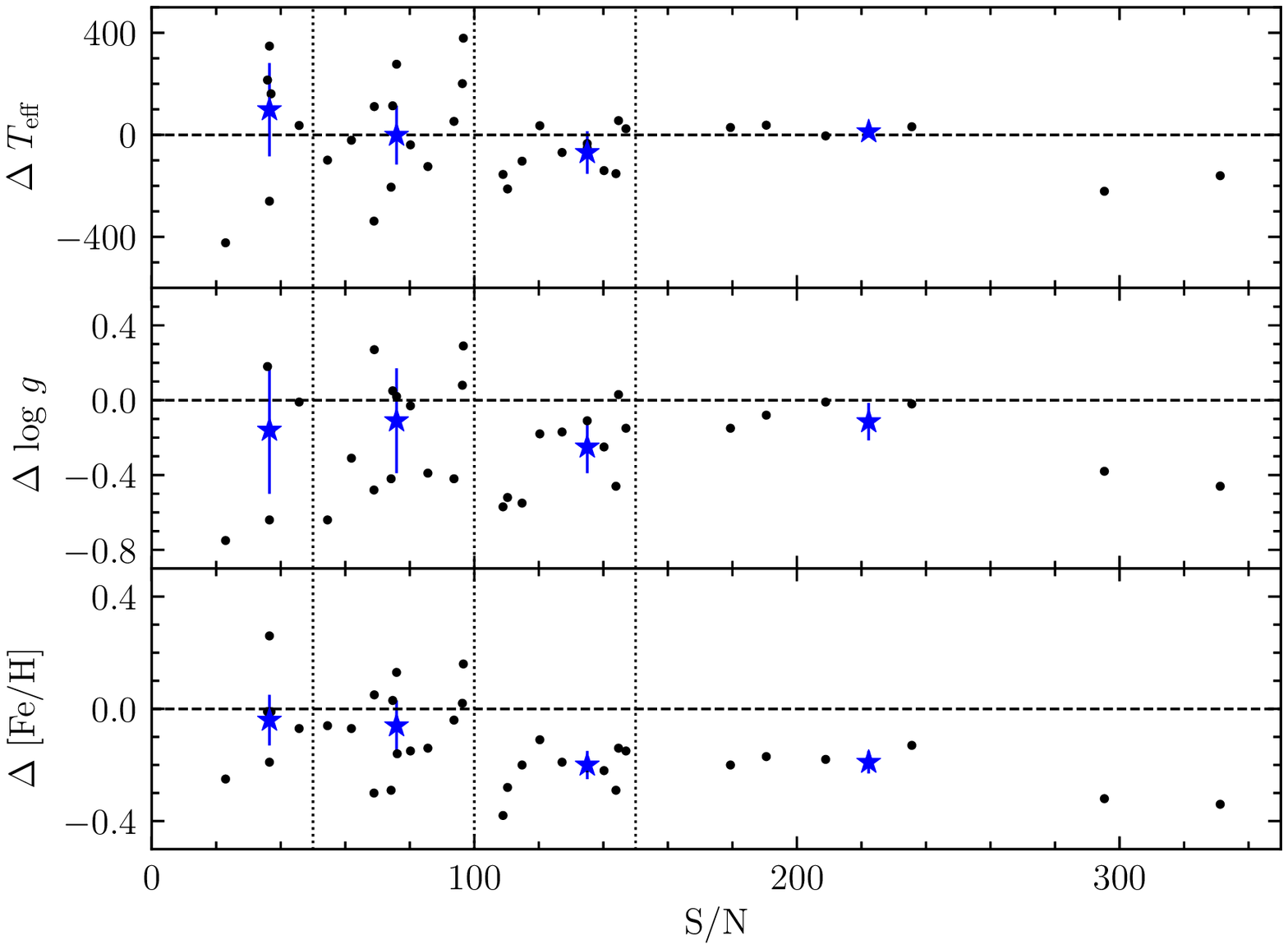}
\caption{Residual plots for the stellar parameters between our estimated values and 
the reference values for our comparison stars, as a function of S/N. The blue star symbol with an error bar indicates 
the median value and its median absolute deviation (MAD) calculated in each S/N range. There 
are four S/N ranges shown ($<$ 50, 50 -- 100, 100 -- 150, and $>$ 150), separated by the vertical dotted lines.}
\label{figure3}
\end{figure}

\begin{figure} [t]
\includegraphics[width=\columnwidth]{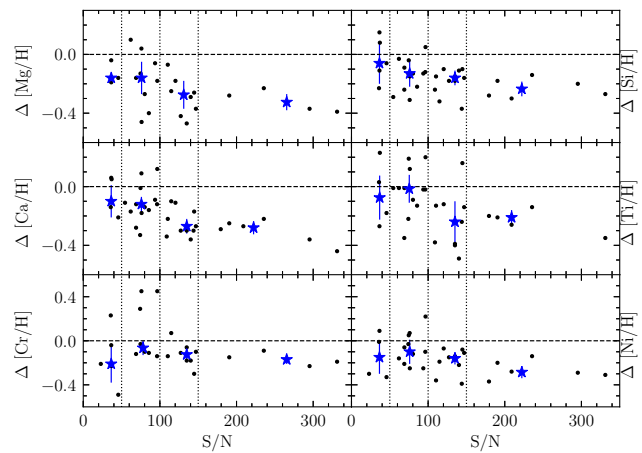}
\caption{Same as in Figure~\ref{figure3}, but for the chemical abundances 
of Mg, Si, Ca, Ti, Cr, and Ni. Note that the lower number of points for 
S/N $<$ 50 results from a lower number of abundance measurements, due to too 
weak absorption features.}
\label{figure4}
\end{figure}

\begin{table*}
\caption{Adopted stellar parameters and chemical abundances of the comparison stars from various literature \label{tab2}}
\centering
\tiny
\begin{tabular}{lcccccccccl}
\toprule
Star       & $T_{\rm eff}$ (K) & $\log~g$ & [Fe/H] & [Mg/H] & [Si/H] & [Ca/H] & [Ti/H] & [Cr/H] & [Ni/H]  & References \\
\midrule
HIP 33582  &     5782          & 4.30     & --0.68 & --0.23 & --0.36 & --0.50 & --0.33 & --0.64 &$\cdots$ & 1, 3, 6, 8, 10   \\
HIP 44075  &     5880          & 4.10     & --0.90 & --0.53 & --0.55 & --0.60 & --0.57 & --0.89 & --0.87  & 1, 2, 3, 4, 5, 7, 11  \\
HIP 59330  &     5749          & 4.02     & --0.75 & --0.45 & --0.42 & --0.58 & --0.47 & --0.74 & --0.75  & 1, 3, 4   \\
HIP 60551  &     5724          & 4.38     & --0.86 & --0.53 & --0.52 & --0.61 & --0.59 & --0.85 & --0.76  & 1, 3, 4  \\
HIP 62882  &     5692          & 3.81     & --1.26 & --0.73 & --0.79 & --0.93 & --0.99 & --1.30 & --1.21  & 1, 3, 5, 11   \\
HIP 64426  &     5892          & 4.19     & --0.76 & --0.39 & --0.50 & --0.62 & --0.58 & --0.81 & --0.77  & 1, 3, 4, 8    \\
HR 2233    &     6240          & 3.97     & --0.19 & --0.06 & --0.11 & --0.11 &  ~0.11 & --0.22 & --0.22  & 4, 5, 6, 7, 11  \\
HR 2721    &     5872          & 4.30     & --0.32 & --0.13 & --0.21 & --0.25 & --0.17 & --0.35 & --0.33  & 4, 5, 7, 9, 11 \\
\bottomrule
\end{tabular}
\tabnote{The numbers in the last column indicate
the following references:
(1) \citet{bensby2003}, (2) \citet{bensby2005},
(3) \citet{fulbright2000}, (4) \citet{lee2011},
(5) \citet{prugniel2011}, (6) \citet{houk2000},
(7) \citet{takeda2005}, (8) \citet{battistini2015},
(9) \citet{mishenina2013}, (10) \citet{delgado2014},
(11) \citet{cenarro2007}. When possible, we took an average of the
listed references for the stellar parameters and elements.}
\end{table*}

\begin{table*}
\caption{Derived stellar parameters and chemical abundances for our program stars \label{tab3}}
\centering
\resizebox{\textwidth}{!}{  
\begin{tabular}{cccccccccccc}
\toprule
SDSS~ID              & Pal14~ID & $T_{\rm eff}$ (K) & $\log~g$ & [Fe/H] & [Mg/Fe] & [Si/Fe] & [Ca/Fe] & [Ti/Fe] & [Cr/Fe] & [Ni/Fe] & [$\alpha$/Fe]   \\
\midrule
J113102.87+665751.1  &  4       & 5038 $\pm$ 235 & 4.47 $\pm$ 0.41 & --0.49 $\pm$ 0.13  &  ~0.48   &  ~0.34   &  ~0.32   &  ~0.37   & --0.07  & --0.11   &  ~0.38  \\
J064337.13+291410.0  &  7       & 5136 $\pm$ 260 & 4.55 $\pm$ 0.48 & --0.60 $\pm$ 0.14  &  ~0.47   &  ~0.21   & --0.20   &  ~0.41   & $\cdots$&  ~0.09   &  ~0.22  \\
J172630.60+075544.0  & 10       & 4986 $\pm$ 404 & 4.77 $\pm$ 0.66 & --0.69 $\pm$ 0.16  & --0.34   &  ~0.40   &  ~0.05   &  ~0.50   & $\cdots$&  ~0.21   &  ~0.15  \\
J095816.39+005224.4  & 15       & 5290 $\pm$ 290 & 4.46 $\pm$ 0.50 & --0.55 $\pm$ 0.15  & --0.13   &  ~0.06   &  ~0.10   & --0.22   &  ~0.62  & --0.10   & --0.05  \\
J074728.84+185520.4  & 16       & 5446 $\pm$ 282 & 4.10 $\pm$ 0.50 &  ~0.27 $\pm$ 0.14  & --0.10   & --0.20   & --0.11   & --0.41   & --0.10  & --0.06   & --0.20  \\
J064257.02+371604.2  & 17       & 5261 $\pm$ 248 & 4.45 $\pm$ 0.42 & --0.12 $\pm$ 0.13  &  ~0.05   & --0.08   &  ~0.05   & --0.09   &  ~0.18  &  ~0.14   & --0.02  \\
\bottomrule
\end{tabular}
}
\tabnote{Note that the systematic offsets are corrected for these values. [$\alpha$/Fe] is a mean value of [Mg/Fe], [Si/Fe], [Ca/Fe], and 
[Ti/Fe]. Typical errors for elements are $\sigma$[Mg/Fe] = 0.27 dex, 
$\sigma$[Si/Fe] = 0.17 dex, $\sigma$[Ca/Fe] = 0.15 dex, $\sigma$[Ti/Fe]
= 0.18 dex, $\sigma$[Cr/Fe] = 0.20 dex, and $\sigma$[Ni/Fe] = 0.18 dex, respectively.}
\end{table*}

\section{Estimation of Stellar Parameters and Chemical Abundances} \label{sec3}

We determined estimates of the stellar parameters ($T_{\rm eff}$,
$\log~g$, and [Fe/H]), and elemental abundances for Mg, Si, Ca, Ti, Cr,
and Ni using the Stellar Parameters And chemical abundances estimator
\citep[SP\_Ace; ][]{boeche2016} code. The SP\_Ace code employs similar
methodology to that used in the RAdial Velocity Experiment \citep[RAVE;
][]{steinmetz2006} chemical abundance pipeline \citep{boeche2011}, which
was developed to derive elemental abundances for the stellar spectra
obtained by the RAVE survey. SP\_Ace estimates the stellar parameters
and abundances based on a library of the equivalent widths (EWs) for
4643 absorption lines. The EWs are generated from a synthetic grid in
the ranges of $3600<T_{\rm{eff}}<7400~\rm{K}$, $0.2<\log~g<5.4$, and
$-2.4<\rm{[M/H]}<0.4$. Each spectrum in the synthetic grid was synthesized 
by MOOG \citep{sneden1973}, after adopting the ALTAS9 model atmospheres \citep{castelli2003}.
Based on input trial values of $T_{\rm{eff}}$,
$\log~g$, and [elements/H], SP\_Ace calculates by interpolation 
the expected EWs using the library to generate a normalized model spectrum. Then,
using the Levenberg-Marquadt method, it attempts to minimize the
$\chi^2$ between the model and the observed spectrum to determine the
stellar parameters and chemical abundances. As the lines 
used to measure the abundances of the individual elements and stellar parameters in 
SP\_Ace are well described in \citet{boeche2016}, we refer the interested 
reader to their paper.

Even though we obtained spectra of both the blue and red channels with
the DIS instrument, we exclusively used the red channel spectra, because
the adopted wavelength ranges (5212 -- 6860\,{\AA} and 8400 -- 8920\,{\AA})
used by SP\_Ace cover a much larger wavelength range in the red. We
applied SP\_Ace to the spectra of the co-added spectra of our program
stars as well as to the spectra of individual exposures of the reference
stars to derive final estimates of the stellar parameters and chemical
abundances.

Before we finalized the stellar parameters and chemical abundances of
our program stars for the analysis of the abundance patterns, we first
compared our derived values with the values for the standard stars from
the various references, as a function of S/N, in order to check for
systematic offsets and estimate the precision of the derived stellar
parameters and chemical abundances. Table~\ref{tab2} lists the adopted 
stellar parameters and chemical abundances of the comparison stars from various 
references.

Figure~\ref{figure3} shows residual plots of $T_{\rm eff}$, $\log~g$,
and [Fe/H] between our values and the reference values, as a function of
S/N, for the standard stars. The blue star symbol with an error
bar indicates the median value and median absolute deviation (MAD),
respectively, in each S/N bin. We considered four bins ($<$ 50, 50 -- 100,
100 -- 150, and $>$ 150), separated by the vertical dotted lines in the
figure. For \teff\ and \logg, we do not notice any
significant trend with S/N. In the case of the metallicity,
our derived value increases with decreasing S/N, as expected. As the typical S/N of
the spectra of our program stars is less than 50, we checked the
systematic offset of each parameter in the range of S/N $<$ 50. We found
a median offset of $T_{\rm eff}$ = 99 $\pm$ 183 K, $\log~g$ = --0.16
$\pm$ 0.34 dex, and [Fe/H] = --0.04 $\pm$ 0.09 dex, respectively, in the sense 
that our values are higher for \teff\ and lower for \logg\ and \feh. The 
uncertainty is the MAD. We decided to adjust the parameter scales by these offsets 
for derivation of the final estimates for our program stars. 

We carried out a similar exercise for the chemical abundances (Mg, Si,
Ca, Ti, Cr, and Ni); Figure~\ref{figure4} displays the results.
Similar to [Fe/H], there is a tendency for larger derived abundances
with decreasing S/R. For S/N $<$ 50, we found median offsets with  of [Mg/H] = --0.16 $\pm$
0.03 dex, [Si/H] = --0.06 $\pm$ 0.14 dex, [Ca/H] = --0.10 $\pm$ 0.11 dex, [Ti/H] =
--0.08 $\pm$ 0.15 dex, [Cr/H] = --0.21 $\pm$ 0.17 dex, and [Ni/H] = --0.15 $\pm$
0.15 dex. The uncertainty is the MAD. We applied these offsets to our program stars.
Table~\ref{tab3} lists the offset-corrected stellar parameters and
chemical abundances for our program stars. The value of [$\alpha$/Fe]
is a mean of the four ratios [Mg/Fe], [Si/Fe], [Ca/Fe], and [Ti/Fe]. The
typical errors on the abundances for our target stars are $\sigma$[Mg/Fe]
= 0.27 dex, $\sigma$[Si/Fe] = 0.17 dex, $\sigma$[Ca/Fe] = 0.15 dex, $\sigma$[Ti/Fe]
= 0.18 dex, $\sigma$[Cr/Fe] = 0.20 dex, and $\sigma$[Ni/Fe] = 0.18 dex. 
These uncertainties and the uncertainties on \teff, \logg, and \feh\ are calculated 
by adding in quadrature the internal error from SP\_Ace and the MAD in the range of 
S/N $<$ 50 for the standard stars. We use these corrected values for the analysis of chemical 
properties throughout this paper. 

As setting S/N values to different limits in Figures \ref{figure3} and 
\ref{figure4} can result in different levels of the offsets in each parameter and abundance, 
we carried out an exercise to evaluate its effect by applying different limits of 
S/N, for example bins of S/N $<$ 50, S/N = 50 -- 90, S/N = 90 -- 150, and S/N $>$ 150. 
We did not find any significant difference from our original S/N limits, but the offsets all 
agreed within the errors. In particular, the differences in the chemical abundances from 
various different S/N limits are less than 0.02 dex, implying that these effect do not 
affect interpretation of the chemical properties of our program stars. 
The reason for the small offsets among the different S/N bins is that we take a 
median value of the several points in each SNR bin. 

Also note that the reason for the smaller offset at the low S/N 
in Figures \ref{figure3} and \ref{figure4} is that, as the noise in a low S/N spectrum 
can mimic extra absorption, the estimated abundance tends to be 
larger. We generally see this tendency in Figures \ref{figure3} and \ref{figure4} 
for our standard stars. For this reason, we adjusted the 
stellar parameters and chemical abundances of our programs by the offsets derived 
from S/N $<$ 50.

\begin{table*}
\caption{Proper motions of our program stars from different catalogs \label{tab4}}
\centering
\resizebox{\textwidth}{!}{
\begin{tabular}{cc|cc|cc|cc}
\toprule
SDSS~ID             & Pal14~ID & \multicolumn{2}{c}{SDSS} & \multicolumn{2}{c}{Ziegerer et al. (2015)} & \multicolumn{2}{c}{{\it Gaia}~DR2}  \\
                    &          &  $\mu_{\alpha}$cos$\delta$ & $\mu_{\delta}$  &  $\mu_{\alpha}$cos$\delta$ & $\mu_{\delta}$ &  $\mu_{\alpha}$cos$\delta$ & $\mu_{\delta}$ \\
                    &          & (mas~yr$^{-1}$) & (mas~yr$^{-1}$) & (mas~yr$^{-1}$) & (mas~yr$^{-1}$) & (mas~yr$^{-1}$) & (mas~yr$^{-1}$) \\
\midrule
J113102.87+665751.1 & 4        & --117.2 $\pm$ 5.9  & 206.8 $\pm$ 5.9 &--12.9 $\pm$ 2.4 & --20.0 $\pm$ 4.1 &--18.01 $\pm$ 0.07  &--26.86 $\pm$ 0.07    \\
J064337.13+291410.0 & 7        & ~--23.6 $\pm$ 2.6  & ~29.9 $\pm$ 2.6 &   $\cdots$      &   $\cdots$       &  ~0.35 $\pm$ 0.38  & --2.43 $\pm$ 0.33    \\
J172630.60+075544.0 & 10       & ~~~19.7 $\pm$ 2.9  &--56.4 $\pm$ 2.9 &   $\cdots$      &   $\cdots$       & --6.21 $\pm$ 0.51  &  ~1.56 $\pm$ 0.48    \\
J095816.39+005224.4 & 15       & ~--58.6 $\pm$ 5.4  &~~~8.1 $\pm$ 5.4 & ~~1.1 $\pm$ 2.2 & ~--2.2 $\pm$ 2.3 & --4.20 $\pm$ 0.57  & --1.27 $\pm$ 0.85    \\
J074728.84+185520.4 & 16       & ~~~~0.8 $\pm$ 5.7  &--58.1 $\pm$ 5.7 & --5.2 $\pm$ 4.6 &  ~~1.1 $\pm$ 5.1 & --3.07 $\pm$ 0.29  & --0.91 $\pm$ 0.24    \\
J064257.02+371604.2 & 17       & ~~~25.2 $\pm$ 2.5  & ~42.1 $\pm$ 2.5 & ~12.5 $\pm$ 3.2 &  ~13.0 $\pm$ 3.5 &  ~1.40 $\pm$ 0.20  & --2.74 $\pm$ 0.19    \\
\bottomrule
\end{tabular}
}
\end{table*}

\begin{table*}
\caption{Velocities and orbital parameters of our program stars \label{tab5}}
\centering
\resizebox{\textwidth}{!}{ 
\begin{tabular}{cccccccccccc}
\toprule
SDSS~ID &  Pal14~ID   & $v_{\rm r}$  & $d$    & $U$     & $V$   & $W$   & $r_{\rm min}$ & $r_{\rm max}$ & $Z_{\rm max}$ & $e$  & $V_{\rm GRF}$  \\ 
       &            & (km\,s$^{-1}$)  & (kpc) & (km\,s$^{-1}$) & (km\,s$^{-1}$) & (km\,s$^{-1}$) & (kpc)  & (kpc) & (kpc) &      & (km\,s$^{-1}$) \\
\midrule
J113102.87+665751.1 &  4       & --65.4 $\pm$ 3.2 & 1.14 &   ~~24.6 $\pm$ 13.1 & --179.9 $\pm$ 30.2 &   ~~14.1  $\pm$ 11.5  &  ~0.90 $\pm$ 0.65   &   ~8.71 $\pm$ 0.20  &  1.04 $\pm$ 0.26 &  0.83 $\pm$ 0.12 & ~49.1 $\pm$ 17.8 \\
J064337.13+291410.0 &  7       &  ~~~9.6 $\pm$ 3.4 & 3.16 &  ~--2.2 $\pm$ 3.6  & ~--32.1 $\pm$  8.5  &  ~--1.7 $\pm$ 5.4   &  ~8.56 $\pm$ 0.46   &   11.11 $\pm$ 0.64  &  0.65 $\pm$ 0.14 &  0.13 $\pm$ 0.04 & 187.9 $\pm$ 8.4  \\
J172630.60+075544.0 & 10       & --37.6 $\pm$ 3.9 & 3.35 &   ~42.0 $\pm$ 6.8  & ~--51.1 $\pm$  9.5  &   ~~84.4 $\pm$ 19.6  &  ~3.77 $\pm$ 0.54   &   ~7.02 $\pm$ 0.21  &  2.53 $\pm$ 0.57 &  0.41 $\pm$ 0.07 & 193.4 $\pm$ 7.9  \\
J095816.39+005224.4 & 15       &  ~~3.2 $\pm$ 3.4 & 2.19 &   ~18.9 $\pm$ 8.4  & ~--13.8 $\pm$  7.4  &  --23.7 $\pm$ 8.3   &  ~8.36 $\pm$ 0.49   &   ~9.52 $\pm$ 0.34  &  1.67 $\pm$ 0.39 &  0.06 $\pm$ 0.03 & 208.4 $\pm$ 6.7  \\
J074728.84+185520.4 & 16       &  ~35.5 $\pm$ 3.0 & 3.71 &   ~41.3 $\pm$ 5.2  & ~~--9.8 $\pm$  3.7  &  --33.2 $\pm$ 9.9   &  10.77 $\pm$ 0.71   &   12.70 $\pm$ 1.20  &  2.05 $\pm$ 0.67 &  0.06 $\pm$ 0.02 & 216.7 $\pm$ 4.6  \\
J064257.02+371604.2 & 17       &  ~--1.7 $\pm$ 3.2 & 1.88 &  --11.3 $\pm$ 2.9  & ~--23.4 $\pm$  6.3  &   ~~8.1 $\pm$ 1.8   &  ~8.20 $\pm$ 0.18   &   ~9.83 $\pm$ 0.37  &  0.56 $\pm$ 0.11 &  0.10 $\pm$ 0.03 & 197.1 $\pm$ 6.2  \\

\bottomrule
\end{tabular}
}
\tabnote{$v_{\rm r}$ is the radial velocity. $U$, $V$, and $W$ velocities were calculated assuming 
$V_{\rm{LSR}}~=~220~\rm{km\,s^{-1}}$ and (--10.1, 4.0, 6.7) km s$^{-1}$ of solar peculiar motion \citep{hogg2005}. 
$V_{\rm GRF}$ is the Galactic rest-frame velocity.}
\end{table*}


\section{Kinematic and Chemical Properties of Our Program Stars} \label{sec4}
\subsection{Kinematic Properties} \label{sec41}

The contradicting results on the status of the SEGUE low-mass candidates
HVSs between Pal14 and \citet{ziegerer2015} stem from the proper motions
that they adopted. Pal14 used the SDSS proper motions \citep{munn2004,
munn2008}, while \citet{ziegerer2015} derived the proper motions with
the images from SDSS, DSS, and UKIDSS. Table~\ref{tab4} summarizes 
these different sets of proper motions for our target stars,
along with those reported in \gaia\ DR2.

Comparison of the three sets of the proper motions reveals that the
proper motions adopted by Pal14 are consistently larger than the other
two; the proper motions of the four stars that \citet{ziegerer2015}
re-derived are closer to the \gaia\ proper motions. Munn (priv.
communication) notes that the reported proper motions from Pal14, based
on SDSS data, are almost certainly the result of mis-identification of
the targets with other nearby stars. The SDSS proper motions of the
other two stars (Pal14 IDs 7 and 10) that \citet{ziegerer2015} could 
not measure the proper motions for are also
larger than those from \gaia\ DR2. These smaller proper motions imply
that our objects are probably bound to the MW. One program star, Pal14
ID 4, exhibits a larger proper motion than the other program stars, and
deserves further investigation.

\begin{figure} [t]
\includegraphics[width=\columnwidth]{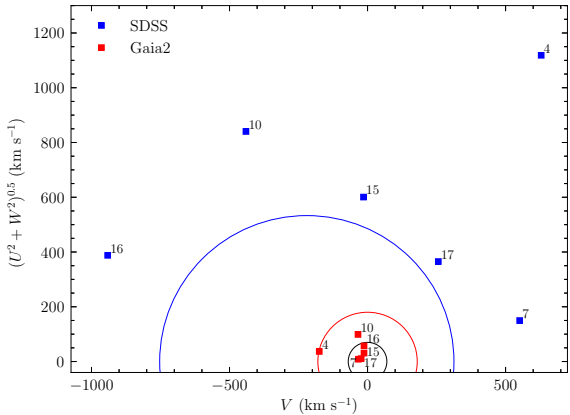}
\caption{Toomre diagram for our sample of stars. The blue symbols indicate 
the velocities calculated with the SDSS proper motions, while the red symbols 
are based on the {\it Gaia} DR2 proper motions. The black and red
circles roughly represent the kinematic 
boundaries of the thin and thick disks, at a constant velocity of 70 km s$^{-1}$ and 
180 km s$^{-1}$ \citep{venn2004}, respectively. The blue line indicates
the local Galactic escape 
speed of $V_{\rm{esc}}$ = 533 km s$^{-1}$ \citep{piffl2014}. The numbers besides each star are 
the sample ID used by Pal14.}
\label{figure5}
\end{figure}

We computed space velocity components and orbital parameters,
as listed in Table~\ref{tab5}, using radial velocities measured from the
obtained spectra, proper motions from {\it Gaia} DR2, and distances
estimated by SEGUE Stellar Parameter Pipeline \citep[SSPP;][]{allende2008,
lee2008a,lee2008b}, based on the methodology of \citet{beers00,beers12}.
The quoted distance uncertainty is on the order of 15--20\%. 
Because the parallaxes to determine the distance do not exist in {\it Gaia} DR2 
for two objects of our program stars, we adopted the photometric distances from the
SSPP, even though the distance uncertainties of four program stars derived by {\it Gaia} DR2 
are much smaller (around 10\%). We also compared the photometric distances 
of our four programs with those from the \gaia\ DR2 parallaxes, and confirmed 
that our photometric distances agreed with the \gaia\ distances within the 
error ranges.

\begin{figure*}[t]
\includegraphics[width=\textwidth]{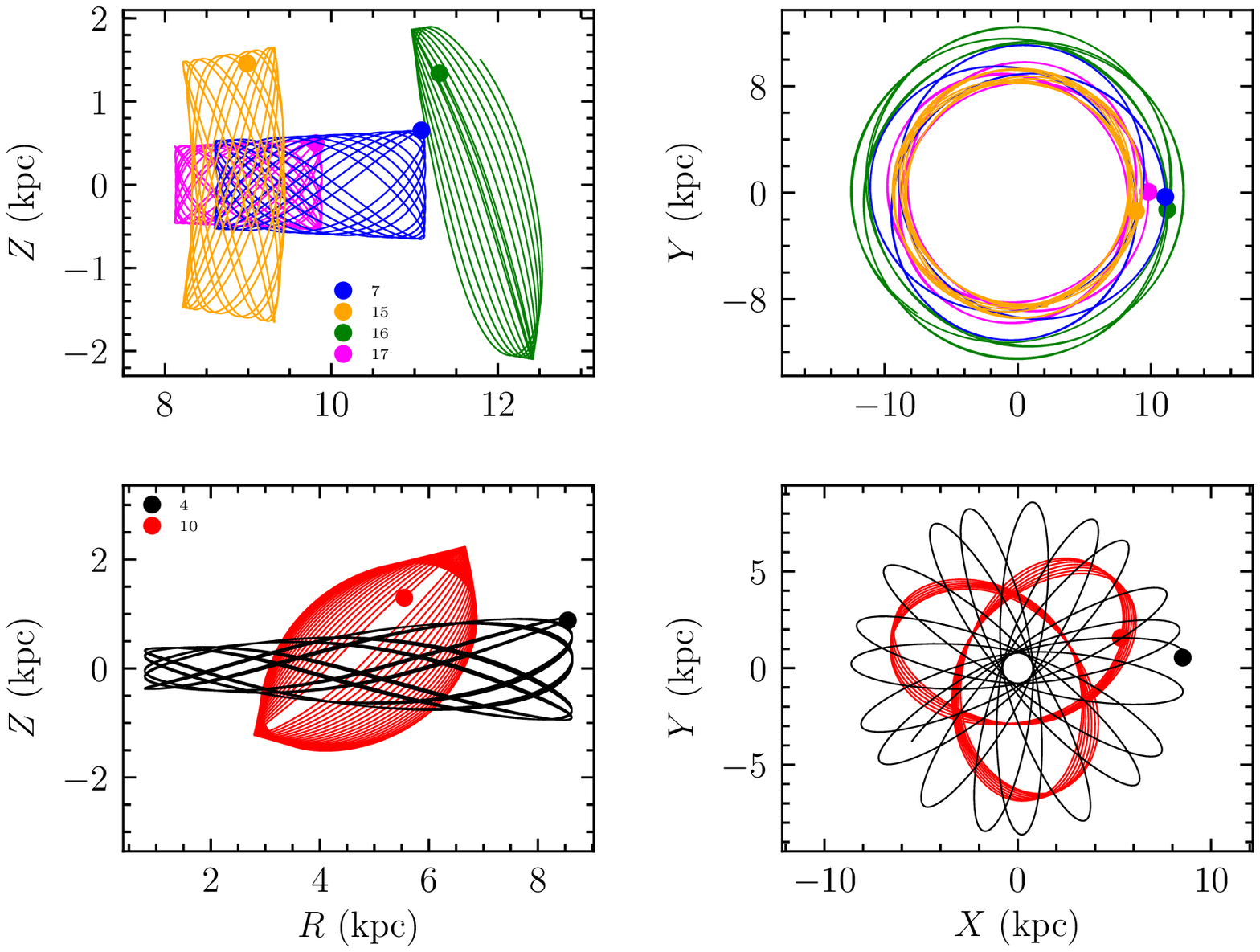}
\caption{Projected orbits for our program stars over 2 Gyr from the
present, in the planes of 
$Z$ and $R$ (left panels) and $X$ and $Y$ (right panels). $Z$ is the
distance from the Galactic plane, while $R$ is the distance from the
Galactic center projected onto the Galactic plane. $X$ and $Y$ are based
on the Cartesian reference system, in which the center of the Galaxy is
at the location at (0, 0) kpc, and the Sun is located at ($X$, $Z$)
=(8.0, 0.0) kpc. The filled circle indicates the current location of
each star.}
\label{figure6}
\end{figure*}

The $U$, $V$, and $W$ velocity components were calculated assuming 220
km s$^{-1}$ of the rotation velocity of the local standard of rest (LSR)
and ($U_\odot$,$V_\odot$,$W_\odot$) = (--10.1, 4.0, 6.7) \kms\ of the
solar peculiar motion \citep{hogg2005}. Each velocity component is
positive in the radially outward direction from the Galactic center for
$U$, the direction of Solar rotation for $V$, and the direction of
the North Galactic Pole for $W$. 

The orbital parameters for individual objects were computed under the
Galactic gravitational potential \emph{MWPotential2014}
\citep{bovy2015}, which is composed of a bulge with a power-law density,
a disk parametrized by a Miyamoto-Nagai potential, with mass 6.8 $\times\ 10^{10}\
M_{\odot}$, and a dark-matter Navarro-Frenk-White halo potential. 
The adopted position of the Sun from the Galactic center is
$R_{\odot}$ = 8.0 kpc. Among the orbital quantities, we derived the
minimum ($r_{\rm min}$) and maximum ($r_{\rm max}$) distances from the
Galactic center, and the maximum distance ($Z_{\rm max}$) from the
Galactic plane during the orbit of a given star. Additionally, the eccentricity
($e$) was obtained from $e =$ ($r_{\rm max}-r_{\rm min}$)/($r_{\rm
max}+r_{\rm min}$). Table~\ref{tab5} lists the calculated
velocity components and the derived orbital parameters. The error on each
velocity and orbital parameter in the table was estimated from the
standard deviation of the distribution of a sample of stars randomly
resampled 100 times, assuming a normal error distribution for the
distance, radial velocity, and proper motion.

Figure~\ref{figure5} is a Toomre diagram for our program stars, which
can be used to kinematically classify different Galactic stellar
components. In the figure, the blue and red squares represent the
velocities computed with the proper motions from SDSS and {\it Gaia}
DR2, respectively. Black and red circles roughly delineate the
boundaries of the thin and thick disks, at constant velocities of 70 km
s$^{-1}$ and 180 km s$^{-1}$ \citep{venn2004}, respectively. The local
Galactic escape speed of $V_{\rm{esc}}$ = 533 \kms\ \citep{piffl2014} is
plotted as a blue circle. The figure clearly indicates that our
spectroscopically-observed HVS candidates are all bound to the MW when
the {\it Gaia} DR2 proper motions are used. According to this diagram,
kinematically, two of our stars appear to belong to the thick disk,
while four of them are likely members of the thin disk.

This component separation is confirmed by the following test.
With the space velocity components ($U$, $V$, $W$) of our program stars 
in hand, we performed a test to compute the likelihood of belonging 
to the thin disk, thick disk, and the halo, following the methodology 
of \citet{bensby2003}. The basic idea is that by assuming that a stellar population 
in the thin disk, thick disk, or the halo has Gaussian distributions 
with different space velocities ($U$, $V$, $W$) and asymmetric drifts, we attempt to separate
our HVS candidates into the thin disk, thick disk, or halo component by 
calculating the probability of belonging to each component. In that process, we 
adopted the local stellar densities, velocity dispersions in $U$, $V$, and $W$, 
and the asymmetric drifts listed in Table 1 of \citet{bensby2003}.

\begin{figure*}[t]
\includegraphics[scale=0.8, width=\textwidth]{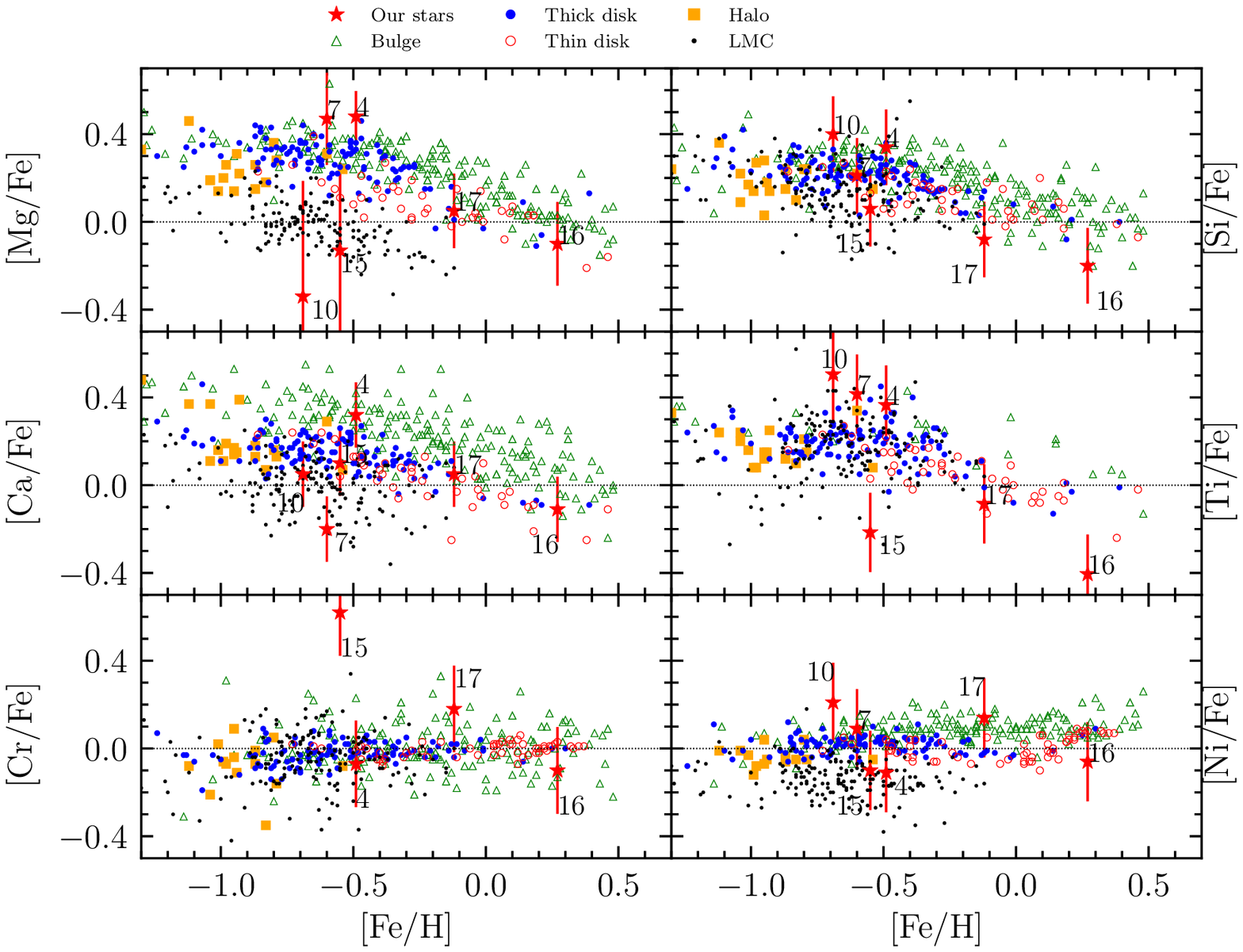}
\caption{Abundance ratios of Mg, Si, Ca, Ti, Cr, and Ni, as a function of
[Fe/H], for stars in the Galactic bulge (green triangles; 
\citealt{alvesbrito2010,johnson2014}), the thick disk (filled-blue circles;
\citealt{alvesbrito2010,bensby2003,reddy2006}), the thin disk (red circles; 
\citealt{alvesbrito2010,bensby2003,reddy2006}), the halo
(filled-orange squares; \citealt{alvesbrito2010, reddy2006}),
and the LMC (black dots; \citealt{swaelmen2013}). 
Our program stars are displayed in red star symbols with the sample ID of Pal14.}
\label{figure7}
\end{figure*}

Based on the computed probability of each program star, 
we derived a relative likelihood of being each component, by comparing 
the probability of being a member of a given component among the three, 
and assigned a star into a component with the higher likelihood. For 
example, if a star has a higher likelihood of being
the thick disk relative to being the thin disk, that is Pr(Thick)/Pr(Thin) $>$ 5, 
this star is assigned the thick disk. On the other hand, the 
thin disk is assigned when a star has Pr(Thick)/Pr(Thin) $<$ 0.5. 

The results of this test for our program stars revealed that only Pal14 IDs 4 and 10 
have Pr(Thick)/Pr(Thin) much larger than 5, and the rest of the stars have less than 0.05.
The probability of being in the halo population is much less, 
Pr(Halo)/Pr(Thick) $<$ 0.01. Only one star with ID 4 has Pr(Halo)/Pr(Thick) $\sim$ 0.9, 
which is still too small to be a halo star. Thus, this exercise proves 
that our program stars belong to the thin or thick disk.

We also compared the Galactic rest-frame velocity ($V_{\rm GRF}$) for
each star with the local escape velocity to check on the probability
that it is bound to the MW. A total of one million Monte Carlo
realizations were carried out to calculate the Galactic rest-frame
velocities after randomly resampling from a normal error distribution of
the radial velocities, distances, and proper motions from {\it Gaia}
DR2. The bound probability is defined by the fraction of the stars that
exceed the local escape velocity to the total number of stars in the
simulation. We found that all our samples are, as expected, bound to the
MW. 

To investigate the orbital characteristics of our program stars, we
integrated the orbit of each star over 2 Gyr from its current position.
Figure~\ref{figure6} shows the projected orbital trajectories of our
program stars into the plane of $Z$ and $R$ (left panels) and $X$ and
$Y$ (right panels). $Z$ is the distance from the Galactic plane, and $R$
is the distance from the Galactic center projected onto the Galactic
plane. $X$ and $Y$ are based on the Cartesian reference system, in which
the center of the Galaxy is at the location at (0, 0) kpc, and the Sun
is located at ($X$, $Z$)=(8.0, 0.0) kpc. The filled circle indicates the
current location of each star. In the figure, we clearly see that four
objects (Pal14 IDs 7, 15, 16, and 17) spend most of their time on orbits
outside the Solar circle (upper-left panel) with nearly circular orbits 
(upper-right panel), with eccentricities less than $e <$ 0.15. Judging from the
orbits and $U$, $V$, $W$ velocities, the Pal14 ID 7 and 17 stars are typical
thin-disk stars, as they are confined to $|Z| < $ 0.7 kpc, while Pal14 ID 15
and 16 stars appear to belong to the thick disk, as they exhibit
excursions above $|Z| >$ 1.5 kpc. 

The bottom panels of Figure~\ref{figure6} indicate that the other two
stars in our sample (Pal14 IDs 4 and 10) are mostly inside the Solar
radius (lower-left panel), with relatively high eccentricity orbits
(lower right); $e$ = 0.83 for Pal14 ID 4 and $e$ = 0.41 for Pal14 ID 10.
Even though Pal14 ID 4 exhibits thick-disk kinematics, an external
origin from a disrupted dwarf galaxy cannot be ruled out for this
object, due to the high eccentricity.

\subsection{Chemical Properties}

Even though the kinematics provide valuable information on which
Galactic component a given star is likely to be a member, the orbits of 
disk stars can change over the course of Galactic evolution due to the perturbations 
by transient spiral patterns or giant molecular clouds. However, since 
the abundance of a chemical element 
for dwarf stars is essentially invariant during its main sequence
lifetime, this can provide additional information on its likely parent
Galactic component, as we explore in this section for our program HVS
candidates.

Among the chemical elements, the so-called $\alpha$-elements such as Mg,
Si, Ca, and Ti, are good indicators of the star-formation history
(duration and intensity) of a stellar population \citep{tinsley1979}.
These elements are produced by successive capture of $\alpha$-particles
in massive stars, which explode as core-collapse supernova (CCSN) that
enrich the surrounding interstellar medium (ISM) with these elements. At
early times, the ISM of a stellar population is enriched by CCSNs,
whereas at later times, by Type Ia SNe, which produce more iron-peak
elements. Consequently, the large enhancements of the
$\alpha$-elements relative to Fe in a stellar population indicates that
it experienced rapid star formation, while the lower values of this
ratio suggests slower, prolonged star formation.

Figure~\ref{figure7} exhibits the distributions of Mg, Si, Ca, Ti, Cr,
and Ni abundances with respect to Fe, as a function of [Fe/H], for four
different Galactic stellar populations -- thin disk (red circle), thick
disk (filled-blue circle), bulge (green triangle), halo (filled-orange
square), and LMC stars (black circles). Our program stars are
represented by star symbols. The chemical abundances of each star in
Galactic stellar populations are adopted from the following references:
bulge stars from \citet{alvesbrito2010,johnson2014}, halo stars from
\citet{alvesbrito2010, reddy2006}, thin and thick disk stars from
\citet{alvesbrito2010,bensby2003,reddy2006}; the LMC stars are from 
\citet{swaelmen2013}. 

From inspection of Figure~\ref{figure7}, the general trends for each
population can be summarized as follows. The thick-disk and halo stars
exhibit similar trends, as they
are rich in $\alpha$-elements, but relatively lower in Cr and Ni.
The thick-disk stars are mostly more metal-rich ([Fe/H] $>$ --0.9)
than the halo stars ([Fe/H] $<$ --0.8). The bulge stars display a wide
range of metallicity, with enhanced $\alpha$-elements in the metallicity
region overlapping with the thick disk. The level of their
$\alpha$-abundances diminishes with increasing metallicity, later
joining the thin disk. However, one distinct pattern is that the Ca
abundances (somewhat true for Ni as well) of the bulge stars are consistently
higher than the thick- and thin-disk populations. Therefore, the Ca
abundance plays a key role in distinguishing a bulge star from a
disk star. The LMC stars exhibit systematically lower abundances for Mg,
Si, Ca, and Ni elements than the other Galactic components, but
overlap with other Galactic stars in Ti and Cr. We included the LMC
stars in the figure because, as claimed by \citet{boubert2016}, there is a
possibility that runaway stars may come from the LMC. We also note that
there is no clear distinction among the Galactic components in the
Fe-peak element Cr. The Ni abundance of bulge stars are
relatively higher in the range of --0.5 $<$ [Fe/H] $<$ 0.2 than any
other Galactic populations. Our program stars, indicated with star
symbols, all have metallicity larger than [Fe/H] = --0.7; none of them
are likely halo stars.  

Comparing the chemical characteristics of each Galactic component with
those of each star in our program sample, while some of our program stars overlap
with one or two Galactic stellar populations, most of our stars exhibit
somewhat deviant abundances from comparison stellar populations. The
Pal14 ID 16 star, which is the most-Fe rich object in our sample, appears to belong to the
thin disk, because the Mg, Si, Ca, and Cr abundances agree with those of
the thin-disk stars, although the Ti and Ni abundances are lower than
the other thin-disk stars. Its chemistry is also overlapped with
that of the bulge population. Pal 14 ID 17 is likely to be associated with the
thin disk, as the level of most of elements is similar to that of the
thin-disk stars, considering the error bars on Cr and Ni. Pal14 ID 4 is likely
to be a bulge star because of the enhancement of all of its
$\alpha$-elements. Pal14 ID 15 appears to be a metal-poor thin-disk star, as its
$\alpha$-element abundances are lower than the thick disk, although the
Cr abundance stands out with respect to the other elements. Pal14 ID 7 may be
a thick-disk star, because its Mg, Si, and Ti are enhanced and the Ca
abundance does not point to the bulge component, but rather the LMC. 
Taking into account the large error bar for Mg, Pal14 ID 10 appears to belong to the
thick disk, with enhanced Si and Ti.  

\begin{table*}
\caption{Possible origin of our program stars \label{tab6}}
\centering
\begin{tabular}{cccc}
\toprule
SDSS~ID &  Pal14~ID   &  Galactic component  & Possible origin(s)   \\

\midrule                        
\midrule                        
J113102.87+665751.1 & 4        & Bulge & Accreted or heated  \\
J064337.13+291410.0 & 7        & Thick &  $\cdots$  \\
J172630.60+075544.0 & 10       & Thick & Runaway or heated \\
J095816.39+005224.4 & 15       & Thin  & Runaway \\
J074728.84+185520.4 & 16       & Thin  & Runaway \\
J064257.02+371604.2 & 17       & Thin  & $\cdots$ \\
\bottomrule
\end{tabular}
\end{table*}

\section{Discussion} \label{sec5}

Even though our HVS stars turned out to be disk stars, 
based on the kinematic probabilistic membership assignment as described in Section~\ref{sec41}, 
some of them have very distinct dynamical properties compared to the canonical 
disk stars, which we discuss below. In what follows, we consider a star 
with high eccentricity to be either an accreted or heated disk star, as stars 
from a disrupted dwarf galaxy are expected to exhibit high eccentricities, and 
dynamical heating mechanism can also produce high eccentricity stars. We regard 
stars with low eccentricity but comparatively large excursions from the Galactic 
plane to be possible runaway stars.

\subsection{SDSS J113102.87$+$665751.1 (Pal14 ID 4)}

As this star has [Fe/H] = --0.49, and a high value of [$\alpha$/Fe] =
0.38, one might consider this object to be a typical thick-disk star.
However, the several individual $\alpha$-elements appear to be more
enhanced with respect to the thick-disk population, and relatively
closer to the bulge population, especially the Ca abundance, which is
the key element to distinguish between the thick-disk and bulge stars.
Thus, it is reasonable to infer that it is a bulge star. 

The calculated orbital parameters for this star are $Z_{\rm max}$ = 1.04
kpc, $e$ = 0.83, $r_{\rm min}$ = 0.90 kpc, and $r_{\rm max}$ = 8.71 kpc;
hence its orbit spans from the near bulge through the Solar radius with
a very high eccentricity. Note that our $r_{\rm min}$ value is rather
smaller than that of \citet{ziegerer2015} due to the slightly larger
proper motions from {\it Gaia} DR2. These chemical and dynamical
characteristics suggest that this star could be born in the bulge, and
expelled to reach the current location by mechanisms such as dynamical
ejection in high stellar density environment or binary supernova
ejection. However, its Galactic rest-frame velocity of $V_{\rm GRF}$ =
49.1 \kms\ is too small to consider that origin likely. Rather, its
orbit points to it being a former member of a disrupted dwarf galaxy or
a dynamically heated disk star.

\subsection{SDSS J064337.13$+$291410.0 (Pal14 ID 7)}

This star has [Fe/H] = --0.60 and [$\alpha$/Fe] = 0.22, properties
associated with a typical thick-disk star. Looking into the individual
$\alpha$-elements, the Mg and Ti abundances are relatively higher and
the Ca abundance is lower than most stars of the thick-disk population.
Only the Si and Ni abundances are close to those of other thick-disk
stars. 

The kinematic characteristics also suggests this object is a thick-disk
star, as the obtained orbital parameters are $Z_{\rm max}$ = 0.65 kpc,
$e$ = 0.13, $r_{\rm min}$ = 8.56 kpc, and $r_{\rm max}$ = 11.11 kpc.
Combining the chemical and kinematic characteristics, this star may be
born in the outer disk, and now located close to the Solar circle.

\subsection{SDSS J172630.60$+$075544.0 (Pal14 ID 10)}

Given a metallicity of [Fe/H] = --0.69 and [$\alpha$/Fe] = 0.15, this
object is also regarded as a likely thick-disk star. Yet, the abundances of
four $\alpha$-elements exhibit a complex pattern; the Ca abundance is in
the thick disk region, whereas the Si and Ti abundances are higher than
the rest of the thick-disk population. The Ni abundance is also larger
than the thick disk. Taking into account the large uncertainty of the Mg
abundance, this object is considered as a thick-disk star. 

Kinematically, however, this object has an eccentricity of $e$ = 0.41
and $r_{\rm min}$ = 3.77 kpc, and its vertical height is as high as
$Z_{\rm max}$ = 2.53 kpc. Combined with the chemical signatures, this
object may be a dynamically heated disk star. It is also possible that
this star may be a disk runaway star, originating from near the bulge,
or a heated disk star, as it reaches a comparatively large vertical
height. The high eccentricity also indicates a possible external origin
from a dissolved dwarf galaxy, but the extent of its orbit (only
exploring 2 kpc away) appears to be too small to be commensurate with an
accreted origin.

\subsection{SDSS J095816.39$+$005224.4 (Pal14 ID 15)}

The metallicity of [Fe/H] = --0.55 and [$\alpha$/Fe] = --0.05 for this
star indicate a metal-poor thin-disk star. The chemical abundances of
individual elements agree with those of the thin-disk population within
the error bars, except the lower Ti and higher Cr abundances than those
of any Galactic population.

The derived orbital parameters of $r_{\rm min}$ = 8.36 kpc, $Z_{\rm
max}$ = 1.67 kpc, and $e$ = 0.06 also suggest a thin-disk star. As this
star reaches up to $Z$ = 1.67 kpc, this is a good candidate for a disk
runaway star ejected by a SNe explosion in a binary system. As
\citet{bromley2009} simulated, some runaway stars exhibit similar
kinematics to the disk or halo stars. 

\subsection{SDSS J074728.84$+$185520.4 (Pal ID 16)}

This object has the highest metallicity in our sample, [Fe/H] = 0.27,
and its [$\alpha$/Fe] value is --0.20. All six elements show
relatively lower abundance ratios than the Sun, which mimic the patterns
of a metal-rich thin-disk star. Hence, one may consider this to belong
to the thin disk.

The orbital parameters of $r_{\rm min}$ = 10.77 kpc, $Z_{\rm max}$ =
2.05 kpc, and $e$ = 0.06 also imply a thin-disk star. As can be seen
from Figure~\ref{figure6}, this object resides in the outer disk and
undergoes large excursions, with a small eccentricity, above the Galactic plane.
Thus, it is more plausible to infer that this object is a runaway star
from the thin-disk population. 

\subsection{SDSS J064257.02$+$371604.2 (Pal ID 17)}

Pal14 ID 17 is a metal-rich ([Fe/H] = --0.12) star with an essentially Solar
alpha ratio ([$\alpha$/Fe] = --0.02). The abundances of the individual
$\alpha$-elements are overlapped with those of the thin-disk population,
whereas the iron-peak elements appear more consistent with the bulge.
Its orbital parameters ($r_{\rm min}$ = 8.20 kpc, $Z_{\rm max}$ = 0.56
kpc, and $e$ = 0.10) also point to a typical thin-disk star.

\section{Summary and Conclusions} \label{sec6}

We have presented a chemodynamical analysis of six low-mass dwarf
stars, alleged to be HVSs by Pal14, in order to determine from which
Galactic component they originate. Based on kinematic analysis using
accurate \gaia\ DR2 proper motions, we confirm that all six objects are
bound to the MW, as noted previously by \citet{ziegerer2015}. Our
conclusion is also upheld by the recent study by \citet{boubert2018},
who performed a detailed investigation of late-type HVS candidates with
proper motions from \gaia\ DR2, and found that almost all known
late-type HVS candidates are bound to the Galaxy. The HVS status for the
low-mass stars identified by Pal14 is mainly due to the incorrect
assignment of the proper motions. 

Nonetheless, we have attempted to characterize the parent Galactic stellar components
and origins of our program stars by taking into account a comparison of
their abundance patterns with various Galactic stellar components and
their orbital properties simultaneously. We note that the kinematic 
probabilistic assignment of their membership to the Galactic component 
has revealed that our program stars belong to the thin or thick disk. However, 
since four of the six stars exhibit distinct dynamical properties and chemical 
characteristics compared to the canonical disk stars, we cannot rule out 
exotic origins such as runaway, disk heating, and accretion from dwarf 
galaxies as discussed in the previous section, and summarize below 
and in Table \ref{tab6}.

We identify two typical disk stars (Pal14 IDs 7 and 17); Pal14 ID 7 is a
typical thick disk star, while Pal ID 17 is a typical thin-disk star, as
these stars have similar abundance patterns to stars of Galactic thick
and thin disks, respectively, and have nearly circular orbits. 

One star (Pal14 ID 4) may originate from the Galactic bulge, as its
orbit passes close to the bulge with a high eccentricity. Moreover, the
abundances of the $\alpha$-elements for this star all agree with those
of other Galactic bulge stars. One may think that this star may be
ejected from the bulge by dynamical ejection or SNe ejection mechanism.
However, its $V_{\rm GRF}$ is too small to consider such an origin
likely. Rather, its high-eccentricity orbit suggests an accreted origin
from a disrupted dwarf galaxy, or dynamical heating.

Pal14 ID 10 may be a runway or heated disk star, as it reaches
as high as 2.53 kpc from the Galactic plane during its orbit, and most
of the chemical abundances within the derived uncertainties are similar
to the thick-disk population.  

Pal14 ID 15 appears to be a runaway from the Galactic disk, because it
exhibits very large excursion (\zmax\ = 1.67 kpc) from the Galactic
plane with small eccentricity ($e$ = 0.06) in an orbit reaching beyond
the Solar radius. The similar chemical abundance pattern to other
metal-poor thin-disk stars support the idea as well. 

Finally, Pal14 ID 16, which has the highest metallicity in our sample,
exhibits similar chemical abundances to a very metal-rich thin-disk star,
while its orbit exhibits a large excursion (\zmax\ = 2.05 kpc) from the
Galactic plane at a location beyond $R$ = 11 kpc. We consider this star
as a runaway star from the Galactic disk.

Although all of our spectroscopically observed candidate HVSs turned out
to be bound to the Galaxy, and not even particularly fast-moving
objects, some of our program stars exhibit exotic orbits. Thus, future
higher resolution spectroscopic follow-up observations for these curious
stars may be of interest, and provide a better understanding of their
origin.

\acknowledgments

We thank anonymous referees for their careful review of this
manuscript, and for pointing out a number of places where we could
improve the clarity of the presentation.

This research has made use of the VizieR catalogue access tool, CDS,
Strasbourg, France. This work has made use of data from the European
Space Agency (ESA) mission {\it
Gaia}\footnote{\url{https://www.cosmos.esa.int/gaia}}, processed by the
{\it Gaia} Data Processing and Analysis Consortium (DPAC)
\footnote{\url{https://www.cosmos.esa.int/web/gaia/dpac/consortium}}.
Funding for the DPAC has been provided by national institutions, in
particular the institutions participating in the {\it Gaia} Multilateral
Agreement.

Funding for SDSS-III was provided by the Alfred P. Sloan Foundation, the
Participating Institutions, the National Science Foundation, and the U.S.
Department of Energy Office of Science. The SDSS-III Web site is \url{http://www.sdss3.org/}.

Y.S.L. acknowledges support from the National Research Foundation (NRF) of
Korea grant funded by the Ministry of Science and ICT (No.2017R1A5A1070354
and NRF-2018R1A2B6003961). T.C.B. acknowledges partial support for this work
from grant PHY 14-30152; Physics Frontier Center/JINA Center for the Evolution
of the Elements (JINA-CEE), awarded by the US National Science
Foundation.


\begin{thebibliography}{}
\bibitem[Abadi et al.(2009)]{abadi2009} Abadi, M.~G., Navarro, J.~F., \& Steinmetz, M.\ 2009, An Alternative Origin for Hypervelocity Stars, ApJL, 691, L63 
\bibitem[Allende Prieto et al.(2008)]{allende2008} Allende Prieto, C., Sivarani, T., Beers, T.~C., et al.\ 2008, The SEGUE Stellar Parameter Pipeline. III. Comparison with High-Resolution Spectroscopy of SDSS/SEGUE Field Stars, AJ, 136, 2070 
\bibitem[Alves-Brito et al.(2010)]{alvesbrito2010} Alves-Brito, A., Mel{\'e}ndez, J., Asplund, M., et al.\ 2010, Chemical similarities between Galactic bulge and local thick disk red giants: O, Na, Mg, Al, Si, Ca, and Ti, A\&A, 513, 35
\bibitem[Battistini \& Bensby (2015)]{battistini2015} Battistini, C., \& Bensby, T. \ 2015, The origin and evolution of the odd-Z iron-peak elements Sc, V, Mn, and Co in the Milky Way stellar disk, A\&A, 577, 9
\bibitem[Beers et al.(2012)]{beers12} Beers, T.~C., Carollo, D., Ivezi{\'c}, {\v Z}., et al.\ 2012, The Case for the Dual Halo of the Milky Way, ApJ, 746, 34 
\bibitem[Beers et al.(2000)]{beers00} Beers, T.~C., Chiba, M., Yoshii, Y., et al.\ 2000, Kinematics of Metal-poor Stars in the Galaxy. II. Proper Motions for a Large Nonkinematically Selected Sample, AJ, 119, 2866
\bibitem[Bensby et al.(2003)]{bensby2003} Bensby, T., Feltzing, S., Lundstrom, I. \ 2003, Elemental abundance trends in the Galactic thin and thick disks as traced by nearby F and G dwarf stars, A\&A, 410, 527
\bibitem[Bensby et al.(2005)]{bensby2005} Bensby, T., Feltzing, S., Lundstrom, I. \ 2005, {$\alpha$}-, r-, and s-process element trends in the Galactic thin and thick disks, A\&A, 433, 185
\bibitem[Blaauw (1961)]{blaauw1961} Blaauw, A. \ 1961, On the origin of the O- and B-type stars with high velocities (the "run-away" stars), and some related problems, BAN, 15, 265
\bibitem[Boeche \& Grebel (2016)]{boeche2016} Boeche, C., \& Grebel, E.~K. \ 2016, SP\_Ace: a new code to derive stellar parameters and elemental abundances, A\&A, 587, 2
\bibitem[Boeche et al.(2011)]{boeche2011} Boeche, C., Siebert, A., Williams, M., et al. \ 2011, The RAVE Catalog of Stellar Elemental Abundances: First Data Release, AJ, 142, 193
\bibitem[Boubert \& Evans(2016)]{boubert2016} Boubert, D., \& Evans, N.~W.\ 2016, A Dipole on the Sky: Predictions for Hypervelocity Stars from the Large Magellanic Cloud, ApJL, 825, L6 
\bibitem[Boubert et al.(2018)]{boubert2018} Boubert, D., Guillochon, J., Hawkins, K., et al.\ 2018, Revisiting hypervelocity stars after Gaia DR2, MNRAS, 479, 2789.
\bibitem[Bovy(2015)]{bovy2015} Bovy, J.\ 2015, galpy: A python Library for Galactic Dynamics, ApJs, 216, 29 
\bibitem[Bromley et al.(2009)]{bromley2009} Bromley, B.~C., Kenyon, S.~J., Brown, W.~R., \& Geller, M.~J.\ 2009, Runaway Stars, Hypervelocity Stars, and Radial Velocity Surveys, ApJ, 706, 925 
\bibitem[Brown(2015)]{brown2015AR} Brown, W.~R.\ 2015, Hypervelocity Stars, ARA\&A, 53, 15 
\bibitem[Brown et al.(2015)]{brown2015} Brown, W.~R., Anderson, J., Gnedin, O.~Y., et al.\ 2015, Proper Motions and Trajectories for 16 Extreme Runaway and Hypervelocity Stars, ApJ, 804, 49 
\bibitem[Brown et al.(2014)]{brown2014} Brown, W.~R., Geller, M.~J., \& Kenyon, S.~J.\ 2014, MMT Hypervelocity Star Survey. III. The Complete Survey, ApJ, 787, 89 
\bibitem[Brown et al.(2005)]{brown2005} Brown, W.~R., Geller, M.~J., Kenyon, S.~J., \& Kurtz, M.~J.\ 2005, Discovery of an Unbound Hypervelocity Star in the Milky Way Halo, ApJ, 622, L33
\bibitem[Castelli \& Kurucz(2003)]{castelli2003} Castelli, F., \& Kurucz, R. L. \ 2003, Modelling of Stellar Atmospheres, ed. N. Piskunov, W.W. Weiss, and D.F. Gray (Published on behalf of the IAU by the ASP, 20) 
\bibitem[Cenarro et al.(2007)]{cenarro2007} Cenarro, A.~J., Peletier, R.~F., Sanchez-Blazquez, P., et al. \ 2007, Medium-resolution Isaac Newton Telescope library of empirical spectra - II. The stellar atmospheric parameters, MNRAS, 374, 664
\bibitem[Cui et al.(2012)]{cui2012} Cui, X.-Q., Zhao, Y.-H., Chu, Y.-Q., et al.\ 2012,  The Large Sky Area Multi-Object Fiber Spectroscopic Telescope (LAMOST), Research in Astronomy and Astrophysics, 12, 1197.
\bibitem[Delgado et al.(2014)]{delgado2014} Delgado Mena, E.~G., Israelian, J.~I., Gonzalez Hernandez, S.~G., et al. \ 2014, Li depletion in solar analogues with exoplanets. Extending the sample, A\&A, 562, 92
\bibitem[Freeman \& Bland-Hawthorn(2002)]{freeman2002} Freeman, K., \& Bland-Hawthorn, J.\ 2002, The New Galaxy: Signatures of Its Formation, ARA\&A, 40, 487 
\bibitem[Fulbright (2000)]{fulbright2000} Fulbright, J.~P. \ 2000, Abundances and Kinematics of Field Halo and Disk Stars. I. Observational Data and Abundance Analysis, AJ, 120, 1841
\bibitem[Gaia Collaboration et al.(2018)]{gaia2018} Gaia Collaboration, Brown, A.~G.~A., Vallenari, A., et al.\ 2018, Gaia Data Release 2. Summary of the contents and survey properties, A\&A, 616, A1.
\bibitem[Gies(1987)]{gies1987} Gies, D.~R.\ 1987, The kinematical and binary properties of association and field O stars, ApJS, 64, 545 
\bibitem[Hawkins \& Wyse (2018)]{hawkins2018} Hawkins, K., \& Wyse, R.~F.~G.\ 2018, The fastest travel together: chemical tagging of the fastest stars in Gaia DR2 to the stellar halo, MNRAS, 481, 1028 
\bibitem[Hills (1988)]{hills1988} Hills, J.~G. \ 1988, Hyper-velocity and tidal stars from binaries disrupted by a massive Galactic black hole, Nature, 331, 687
\bibitem[Hogg et al.(2005)]{hogg2005} Hogg, D.~W., Blanton, M.~R., Roweis, S.~T., et al. \ 2005, Modeling Complete Distributions with Incomplete Observations: The Velocity Ellipsoid from Hipparcos Data, ApJ, 629, 268
\bibitem[Houk \& Swift (2000)]{houk2000} Houk, N., \& Swift, C. \ 2000, VizieR Online Data Catalog: Michigan Catalogue of HD stars, Vol.5 (Houk+, 1999), VizieR Online Data Catalog, 3214
\bibitem[Johnson et al.(2014)]{johnson2014} Johnson, C.~I., Rich, R.~M., Kobayashi, C., Kunder, A., \& Koch, A. \ 2014, Light, Alpha, and Fe-peak Element Abundances in the Galactic Bulge, AJ, 148, 67
\bibitem[Kollmeier \& Gould(2007)]{kollmeier2007} Kollmeier, J.~A., \& Gould, A.\ 2007, Where Are the Old-Population Hypervelocity Stars?, ApJ, 664, 343 
\bibitem[Kollmeier et al.(2009)]{kollmeier2009} Kollmeier, J.~A., Gould, A., Knapp, G., \& Beers, T.~C. \ 2009, Old-population Hypervelocity Stars from the Galactic Center: Limits from the Sloan Digital Sky Survey, ApJ, 697, 1543
\bibitem[Kurtz \& Mink (1998)]{kurtz1998} Kurtz, M.~J. \& Mink, D.~J. \ 1998, RVSAO 2.0: Digital Redshifts and Radial Velocities, PASP, 110, 934
\bibitem[Lawrence et al. (2007)]{lawrence2007} Lawrence, A., Warren, S.~J., Almaini, O., et al.\ 2007, The UKIRT Infrared Deep Sky Survey (UKIDSS), MNRAS, 379, 1599 
\bibitem[Lee et al.(2011)]{lee2011} Lee, Y.~S., Beers, T.~C., Prieto, C.~A., et al. \ 2011, The SEGUE Stellar Parameter Pipeline. V. Estimation of Alpha-element Abundance Ratios from Low-resolution SDSS/SEGUE Stellar Spectra, AJ, 141, 90
\bibitem[Lee et al.(2008a)]{lee2008a} Lee, Y.~S., Beers, T.~C., Sivarani, T., et al.\ 2008a, The SEGUE Stellar Parameter Pipeline. I. Description and Comparison of Individual Methods, AJ, 136, 2022 
\bibitem[Lee et al.(2008b)]{lee2008b} Lee, Y.~S., Beers, T.~C., Sivarani, T., et al.\ 2008b, The SEGUE Stellar Parameter Pipeline. II. Validation with Galactic Globular and Open Clusters, AJ, 136, 2050 
\bibitem[Li et al.(2012)]{li2012} Li, Y., Luo, A., Zhao, G., et al. \ 2012, Metal-poor Hypervelocity Star Candidates from the Sloan Digital Sky Survey, ApJL, 744, 24
\bibitem[Li et al.(2015)]{li2015} Li, Y.-B., Luo, A.-L., Zhao, G., et al.\ 2015, 19 low mass hypervelocity star candidates from the first data release of the LAMOST survey, Research in Astronomy and Astrophysics, 15, 1364.
\bibitem[Mishenina et al.(2013)]{mishenina2013} Mishenina, T.~V., Pignatari, M., Korotin, S.~A., Soubiran, C., et al. \ 2013, Abundances of neutron-capture elements in stars of the Galactic disk substructures, A\&A, 552, 128
\bibitem[Munn et al.(2004)]{munn2004} Munn, J.~A., Monet, D.~G., Levine, S.~E., et al.\ 2004, An Improved Proper-Motion Catalog Combining USNO-B and the Sloan Digital Sky Survey, AJ, 127, 3034 
\bibitem[Munn et al.(2008)]{munn2008} Munn, J.~A., Monet, D.~G., Levine, S.~E., et al.\ 2008, Erratum: "an Improved Proper-Motion Catalog Combining Usno-B and the Sloan Digital Sky Survey" (2004, AJ, 127, 3034), AJ, 136, 895 
\bibitem[Palladino et al.(2014)]{palladino2014} Palladino, L.~E., Schlesinger, K.~J., Holley-Bockelmann, K., et al. \ 2014, Hypervelocity Star Candidates in the SEGUE G and K Dwarf Sample, ApJ, 780, 7
\bibitem[Piffl et al.(2014)]{piffl2014} Piffl, T., Scannapieco, C., Binney, J., et al.\ 2014, The RAVE survey: the Galactic escape speed and the mass of the Milky Way, A\&A, 562, A91 
\bibitem[Poveda et al.(1967)]{poveda1967} Poveda, A., Ruiz, J., \& Allen, C. \ 1967, Run-away Stars as the Result of the Gravitational Collapse of Proto-stellar Clusters, Boletin de los Observatorios Tonantzintla y Tacubaya, 4, 86
\bibitem[Prugniel et al.(2011)]{prugniel2011} Prugniel, Ph., Vauglin, I., \& Koleva, M. \ 2011, The atmospheric parameters and spectral interpolator for the MILES stars, A\&A, 531, 165
\bibitem[Reddy et al.(2006)]{reddy2006} Reddy, B.~E., Lambert, D.~L., \& Allende~Prieto, C. \ 2006, Elemental abundance survey of the Galactic thick disc, MNRAS, 367, 1329
\bibitem[Sneden (1973)]{sneden1973} Sneden, C. \ 1973, Carbon and Nitrogen Abundances in Metal-Poor Stars, PhD thesis, Univ. Texas at Austin
\bibitem[Steinmetz et al.(2006)]{steinmetz2006} Steinmetz, M., Zwitter, T., Siebert, A., et al.\ 2006, The Radial Velocity Experiment (RAVE): First Data Release, AJ, 132, 1645 
\bibitem[Stone(1991)]{stone1991} Stone, R.~C.\ 1991, The space frequency and origin of the runaway O and B stars, AJ, 102, 333 
\bibitem[Takeda \& Honda (2005)]{takeda2005} Takeda, Y., \& Honda, S. \ 2005, Photospheric CNO Abundances of Solar-Type Stars, PASJ, 57, 65
\bibitem[Tauris (2015)]{tauris2015} Tauris, T.~M. \ 2015, Maximum speed of hypervelocity stars ejected from binaries, MNRAS, 448, 6
\bibitem[Tauris \& Takens (1998)]{tauris1998} Tauris, T.~M., Takens, R.~J. \ 1998, Runaway velocities of stellar components originating from disrupted binaries via asymmetric supernova explosions, A\&A, 330, 1047
\bibitem[Tetzlaff et al.(2011)]{tetzlaff2011} Tetzlaff, N., Neuh{\"a}user, R., \& Hohle, M.~M.\ 2011, A catalogue of young runaway Hipparcos stars within 3 kpc from the Sun, MNRAS, 410, 190 
\bibitem[Tinsley(1979)]{tinsley1979} Tinsley, B.~M.\ 1979, Stellar lifetimes and abundance ratios in chemical evolution, ApJ, 229, 1046 
\bibitem[Van der Swaelmen et al.(2013)]{swaelmen2013} Van der Swaelmen, M., Hill, V., Primas, F., et al.\ 2013, Chemical abundances in LMC stellar populations. II. The bar sample, A\&A, 560, A44.
\bibitem[Venn et al.(2004)]{venn2004} Venn, K.~A., Irwin, M., Shetrone, M.~D., Tout, C.~A., Hill, V., Tolstoy, E. \ 2004, Stellar Chemical Signatures and Hierarchical Galaxy Formation, AJ, 128, 1177
\bibitem[Watkins et al.(2010)]{watkins2010} Watkins, L.~L., Evans, N.~W., \& An, J.~H.\ 2010, The masses of the Milky Way and Andromeda galaxies, MNRAS, 406, 264 
\bibitem[Xue et al.(2008)]{xue2008} Xue, X.~X., Rix, H.~W., Zhao, G., et al.\ 2008, The Milky Way's Circular Velocity Curve to 60 kpc and an Estimate of the Dark Matter Halo Mass from the Kinematics of ~2400 SDSS Blue Horizontal-Branch Stars, ApJ, 684, 1143 
\bibitem[Yanny et al.(2009)] {yanny2009} Yanny, B., Rockosi, C., Newberg, H.~J., et al. \ 2009, SEGUE: A Spectroscopic Survey of 240,000 Stars with g = 14-20, AJ, 137, 4377
\bibitem[York et al.(2000)] {york2000} York, D.~G., Adelman, J., Anderson, J.~E., Jr., et al. \ 2000, The Sloan Digital Sky Survey: Technical Summary, AJ, 120, 1579
\bibitem[Yu \& Tremaine(2003)]{yu2003} Yu, Q., \& Tremaine, S.\ 2003, Ejection of Hypervelocity Stars by the (Binary) Black Hole in the Galactic Center, ApJ, 599, 1129 
\bibitem[Zhong et al.(2014)]{zhong2014} Zhong, J., Chen, L., Liu, C., et al. \ 2014, The Nearest High-velocity Stars Revealed by LAMOST Data Release 1, ApJL, 789, L2
\bibitem[Ziegerer et al.(2015)]{ziegerer2015} Ziegerer, E., Volkert, M., Heber, U., et al.\ 2015, Candidate hypervelocity stars of spectral type G and K revisited, A\&A, 576, 14






\end{thebibliography}
\end{document}